\documentclass[%
 reprint,
 amsmath,amssymb,
 aps,
prb,
]{revtex4-1}

\usepackage{graphicx}% Include figure files
\usepackage{dcolumn}% Align table columns on decimal point
\usepackage{bm}% bold math
\usepackage{xcolor}
\usepackage{stackengine}
\usepackage[normalem]{ulem}

\newcommand {\braket}[2]{ \langle #1 | #2  \rangle }

\newcommand {\ket}[1]{| #1 \rangle}
\newcommand {\bra}[1]{\langle  #1 | }

\def\PARAMS{{\rm PARAMS}}
\def\COST{{\rm COST}}

\def\br{\pmb{r}}

\begin{document}

%\preprint{APS/123-QED}

\title{Atomic cluster expansion and wave function representations}

\author{Ralf Drautz} 
\affiliation{ICAMS, Ruhr-Universit\"at Bochum, Bochum, Germany}
\email[]{ralf.drautz@rub.de}
\author{Christoph Ortner}
\affiliation{Department of Mathematics, University of British Columbia, Vancouver, BC, Canada V6T 1Z2}

\date{\today}% It is always \today, today,

\begin{abstract}
    The atomic cluster expansion (ACE) has been highly successful for the parameterisation of symmetric (invariant or equivariant) properties of many-particle systems. Here, we generalize its derivation to anti-symmetric functions. We show how numerous well-known linear representations of wave functions naturally arise within this framework and we explore how recent successful nonlinear parameterisations can be further enhanced by employing ACE methodology. From this analysis we propose a wide design space of promising wave function representations.
\end{abstract}

\pacs{Valid PACS appear here}% PACS, the Physics and Astronomy
\maketitle

%\tableofcontents

\section{Introduction\label{sec:Intro}}

The atomic cluster expansion (ACE) was recently introduced as a polynomial expansion for the parameterization of the interatomic interaction \cite{Drautz19} that explicitly encodes permutation and rotation symmetries. It is a {\em complete} representation\cite{Dusson20}, which means that any conceivable form of the interaction between atoms can in principle be represented. The framework has been extended to include magnetism and charge transfer, to handle vectorial and tensorial objects \cite{Drautz20} as well as message passing networks \cite{Bochkarev2022_multilayer,Batatia2022_Design}. Efficient implementations \cite{lysogorskiy_performant_2021} and parameterization software \cite{bochkarev_efficient_2022, ACEsuit, fitSNAP} are available. Previously we provided ample evidence that ACE models can compete with and oftentimes outperform far more complex and computationally expensive deep learning models. 

Here we explore from a theoretical perspective the
deep connections between symmetric and anti-symmetric functions in the context of the ACE framework and how to employ this framework to discretize many-electron wave functions. First, we demonstrate how a number of well-known {\em linear} parameterisations naturally arise from a polynomial expansion by adapting the ACE framework. In particular, we propose a single-particle basis that can explicitly or implicitly incorporate electron-electron and electron-nuclei cusps to accelerate the convergence of the discretisation.

An important motivation for this work is the remarkable success of recent machine learning representations of wave functions\cite{Hermann2020-bf, Pfau2020-zv, han2020solving, Spencer21, Glielmo2020-ls, Rath2020-xr, Luo2019-lb, Carleo18, Cai2018-fw, Hibat-Allah2020-ex, Szabo2020-as}. 
Therefore, we also review the most important nonlinear parameterisations (Slater-Jastrow, Backflow) and discuss how ACE representations of {\em symmetric} functions provide building blocks towards alternative and/or generalized parameterisations of those representations.  
A potential advantage of our framework is a clear interpretation of natural approximation parameters and a flexible choice of those parameters, which will in particular enable the study of convergence, both theoretically and numerically. In this way we contribute to a systematic interpolation between classical and modern nonlinear parameterisations of wave functions and a systematic exploration of this design space.

\section{Cluster expansion \label{sec:ACE}}
The state of a particle is described by a variable $x$, which can include the position and several other features of the particle. For atomistic simulations one chooses $x = (\br, \mu)$, where $\br$ is the position and $\mu$ the chemical species, whereas electrons are described by $x = (\br, \sigma)$ with $\sigma$ the spin state.
We start from an orthonormal and complete set of one-particle basis functions, $\phi_v(x)$, 
\begin{align}
\braket{v}{u} = \sum_{\sigma} \int  d\pmb{r} \, \phi_{v}^*(x) \phi_{u}(x)  &= \delta_{v u} \,, \label{eq:orth1} \\
\sum_v \ket{v} \bra{v} = \sum_v \phi_{v}^*(x) \phi_{v}(x')  &=  \delta(x - x')
% \delta_{\sigma \sigma'} 
\,,\label{eq:comp1}
\end{align}
with basis function indices $v = 0, 1, 2, \dots$. Leaving away any symmetry considerations for the moment, it is straightforward to write down a general expansion of a $N$-particle function $\Psi(1,2, \dots, N)$ as 
\begin{align}
 &\Psi = \sum_v \sum_{i}^{N} c_{v}^i  \phi_{v}(i) + \sum_{v_1 v_2} \sum_{i_1 i_2}^N c_{v_1 v_2}^{i_1 i_2} \phi_{v_1}(i_1) \phi_{v_2}(i_2) \, + \nonumber \\
       & \dots + \sum_{v_1 \dots v_N  } \sum_{i_1 \dots i_N}^N c_{v_1 v_2 \dots v_N}^{i_1 i_2 \dots i_N} \phi_{v_1}(i_1) \phi_{v_2}(i_2) \dots \phi_{v_N}(i_N) \,, \label{eq:genACE}
\end{align}
where $\phi_{v}(i)= \phi_{v}(x_i)$, $\phi_{0}(i) = 1$ and the summation over basis function indices is limited to $v_i >0$.  (Until we impose symmetries, the coefficients $c_{v_1\dots v_n}^{i_1 \dots i_n}$ must carry the labels $i_1, \dots i_n$  of the interacting particles in each term.)
%\cco{If $\phi_0 = 1$ then we don't need the sum over 1-body, 2-body etc.; it is included in the order-$N$ group.} \crd{Hopefully this is clear now.} 
The sum over indices $i_1 i_2 \dots$ is unrestricted, i.e. contributions $i_n = i_m$ are also taken into account. 
This expansion is referred to as the atomic cluster expansion \cite{Drautz19,Drautz20}, see also Ref.~\onlinecite{Sanchez84} for an analogous expansion for discrete variables. One would naturally constrain $c_{v_1 v_2 \dots v_N}^{i_1 i_2 \dots i_N} = 0$ if any two particles are identical $i_n = i_m$ to avoid unphysical self-interactions. As the unphysical self-interactions can be removed from the expansion by modified lower order expansion coefficients \cite{Drautz19,Drautz20,Dusson20}, we will not enforce this constraint in the following. 

Because the representation Eq.(\ref{eq:genACE}) is constructed from products of single particle functions, one may choose the expansion coefficients to be symmetric with respect to simultaneous change of the particle indices $i_n \leftrightarrow i_m$ and the corresponding basis indices  $v_n \leftrightarrow v_m$, illustrated here for the $N$-product coefficient,
\begin{equation}
c_{v_1 v_2 \dots v_n \dots v_m \dots v_N}^{i_1 i_2 \dots i_n \dots i_m \dots i_N} = c_{v_1 v_2 \dots v_m \dots v_n \dots v_N}^{i_1 i_2 \dots i_m \dots i_n \dots i_N} \,, \label{eq:sym0}
\end{equation}
without limiting the generality of the expansion and we will assume this symmetry for the following. 
% This just follows as the atomic cluster expansion is constructed from products of single particle functions. 
Then in Eq.(\ref{eq:genACE}) either the sum over particles $i$ or basis function $v$ can be restricted to $i_1 \leq i_2 \leq \dots i_N$ or $v_1 \leq v_2 \leq \dots v_N $, which leads to different representations of symmetric and anti-symmetric functions that will be discussed in the next sections.

\subsection{Symmetric functions}

If the $N$-particle function is symmetric with respect to exchange of particles
\begin{equation}
\begin{aligned}
    & \quad \,\Psi(1,2, \dots,n,\dots,m,\dots, N) \\
    &=  \Psi(1,2, \dots,m,\dots,n,\dots, N) \,,
\end{aligned}
\end{equation}
then the expansion coefficients are symmetric with respect to exchange of particles, which means that the particle index can be dropped from the expansion coefficients and the expansion written as
\begin{align} 
\Psi &= \sum_v \sum_{i}^{N} c_{v}  \phi_{v}(i) + \sum_{v_1 v_2} \sum_{i_1 i_2}^N c_{v_1 v_2} \phi_{v_1}(i_1) \phi_{v_2}(i_2) + \dots \nonumber \\
       &+\!\! \sum_{v_1 v_2 \dots v_N  } \sum_{i_1 i_2 \dots i_N}^N c_{v_1 v_2 \dots v_N} \phi_{v_1}(i_1) \phi_{v_2}(i_2) \dots \phi_{v_N}(i_N) \,.
   \label{eq:sym_naive_expansion}
\end{align}
From Eq.(\ref{eq:sym0}) the expansion coefficients are symmetric with respect to exchange of basis function indices $v_n \leftrightarrow v_m$, that is, 
\begin{equation}
c_{v_1 v_2 \dots v_n \dots v_m \dots v_N} = c_{v_1 v_2 \dots v_m \dots v_n \dots v_N}\,. \label{eq:sym1}
\end{equation}
This enables to carry out the summation over particles $i$ first followed by summation over basis function indices $v$ (sometimes called "density trick") and enables very efficient expansions for atomistic simulations  \cite{Drautz19, Dusson20, Lysogorskiy21}. To this end the atomic basis is introduced
\begin{equation}
A_v = \sum_i \phi_{v}(i) \,, \label{eq:AB}
\end{equation}
and the expansion written as
\begin{align} \label{eq:Psi_sym_final}
\Psi &= \sum_v c_{v}  A_{v} + \sum_{v_1 v_2} c_{v_1 v_2} A_{v_1} A_{v_2} + \dots \nonumber \\
       &+ \sum_{v_1 v_2 \dots v_N} c_{v_1 v_2 \dots v_N} A_{v_1} A_{v_2} \dots A_{v_N} \,.
\end{align}
From Eq.(\ref{eq:sym1}) the summations can be restricted further to ordered $v_1 \leq v_2 \leq \dots$.

As representation of functions in high dimension is plagued by the curse of dimensionality, it is particularly interesting to highlight the improvement in computational cost that the ACE expansion affords over a naive tensor product approximation. First, we point out that through a recursive evaluation of the basis functions $A_{v_1} \cdots A_{v_n}$, the cost, ${\rm COST}$, of evaluating Eq.(\ref{eq:Psi_sym_final}) becomes directly proportional to the number of parameters, ${\rm PARAMS}$; to be precise, ${\rm COST} \sim 2 \times {\rm PARAMS}$. Counting the number of parameters requires more information about the basis choice than we have provided so far. However, even in general it is intuitively clear that in the limit of a complete basis set where most tuples $(v_1\dots v_N)$ will be strictly ordered, the number of parameters is reduced approximately by a factor $N!$. 

We can say more if we make a specific choice, and will focus for simplicity on the case of a total degree approximation which is particularly natural when approximation smooth (analytic) $\Psi$. This results in restricting the sums $\sum_{v_1 v_2 \dots v_n}$ in Eq.(\ref{eq:Psi_sym_final}) over the ordered tuples $(v_1, \dots, v_n)$ to $\sum_{t = 1}^n {\rm deg}(\phi_{v_t}) \leq D$. Here, ${\rm deg}(\phi_v)$ denotes the degree of the basis function $\phi_{v}$, which depends on the choice of basis, but is almost always canonical and $D$ is the total degree.  \citet{BaDuOrt21} make this precise and show that for a $d$-dimensional discretisation any natural choice of basis will lead to 
both asymptotic and pre-asymptotic bounds,
\begin{equation} \label{eq:TD_bound}
        \PARAMS 
        \lesssim 
        \begin{cases} 
            c^N \frac{D^{dN}}{(dN)! N!}, & \text{as $D \to \infty$}, \\ 
            \exp( c D^{1 - 1/d} ), & \text{uniformly in $N, D$.}
        \end{cases}
\end{equation}
The first bound makes precise the intuition that the basis size is reduced by a factor $N!$ compared with an unsymmetrized approximation. The second bound is particularly interesting in two respects: (1) it does not rely on the $D \to \infty$ asymptotics which cannot be practically reached; (2) It is remains super-algebraic but 
independent of the number of particles $N$, which results in the powerful suggestion that approximation quality may be independent of dimensionality.

\subsection{Anti-symmetric functions \label{sec:anti}}

The expansion may also be simplified significantly for anti-symmetric $N$-particle functions.\footnote{For simplicity of presentation we do not discuss spins explicitly, but note that they can be treated entirely analogously as chemical species in the original ACE models\cite{Drautz19,Dusson20}, replacing a position variable ${\bm x} = {\bm r}$ with a position-spin pair ${\bm x} = ({\bm r}, \sigma)$.} Then,
\begin{equation}
\begin{split}
    & \Psi(1,2, \dots,n,\dots,m,\dots, N)  \\ 
    =\,&  - \Psi(1,2, \dots,m,\dots,n,\dots, N) \,.
\end{split}
\end{equation}
As the exchange of two particles must change the sign of the function, it is clear from the general expansion Eq.(\ref{eq:genACE}) that only the $N$-product may contribute. The expansion coefficient of the $N$-product has to fulfill
\begin{equation}
\begin{split}
c_{v_1 v_2 \dots v_n \dots v_m \dots v_N}^{i_1 i_2 \dots i_n \dots i_m \dots i_N} 
&= - c_{v_1 v_2 \dots v_n \dots v_m \dots v_N}^{i_1 i_2 \dots i_m \dots i_n \dots i_N} \,, \\
&= - c_{v_1 v_2 \dots v_m \dots v_n \dots v_N}^{i_1 i_2 \dots i_n \dots i_m \dots i_N} \,,
\end{split}
\end{equation}
where the second identity follows from Eq.(\ref{eq:sym0}). 
This allows to factorize the expansion coefficient as
\begin{equation}
c_{v_1 v_2 \dots v_N}^{i_1 i_2 \dots i_N} = \epsilon_{i_1 i_2 \dots i_N} \epsilon_{v_1 v_2 \dots v_N} c_{v_1 v_2 \dots v_N} \,,
\end{equation}
with $\epsilon_{i_1 i_2 \dots i_N} = 1$ for an even particle permutation, $\epsilon_{i_1 i_2 \dots i_N} = -1$ for an odd permutation, $\epsilon_{i_1 i_2 \dots i_N} = 0$ if two or more indices are identical $i_n = i_m$, and equivalently for $\epsilon_{v_1 v_2 \dots v_N}$, and where the expansion coefficients are symmetric
\begin{equation}
c_{v_1 v_2 \dots v_n \dots v_m \dots v_N} = c_{v_1 v_2 \dots v_m \dots v_n \dots v_N}\,.
\end{equation}
As for the general expansion Eq.(\ref{eq:genACE}) and using Eq.(\ref{eq:sym0}), one may limit the sums either to ordered particle indices $i_1 < i_2 < \dots < i_N$ or ordered basis function indices $v_1 < v_2 < \dots < v_N$. This implies that one has some freedom for anti-symmetric expansions that we discuss next. 

\subsubsection{Restricted basis function summation}

We first discuss anti-symmetric expansions when the summation over basis functions is limited $v_1 < v_2 < \dots < v_n$, while the particle indices are completely summed over.

\paragraph{Configuration interaction}

With $v_1 < v_2 < \dots < v_N$, Eq.(\ref{eq:genACE}) reduces to a general anti-symmetric expansion given by
\begin{equation} 
\Psi =  \sum_{v_1 < v_2 < \dots < v_N  } c_{v_1 v_2 \dots v_N}  D_{v_1 v_2 \dots v_N} \,, \label{eq:asymACEeps}
\end{equation}
with the Slater determinant
\begin{align} \label{eq:standard_slater_det}
    &D_{v_1 v_2 \dots v_N}  \\ 
    \notag 
    &=  \frac{1}{\sqrt{N!}} \sum_{i_1 i_2 \dots i_N}^N \epsilon_{i_1 i_2 \dots i_N}  \phi_{v_1}(i_1) \phi_{v_2}(i_2) \dots \phi_{v_N}(i_N) .
\end{align}    
The Slater determinant takes the same role as the product basis $A_{v_1} A_{v_2} \cdots A_{v_N}$ in Eq.~\eqref{eq:Psi_sym_final} for symmetric functions. The linear expansion Eq.(\ref{eq:asymACEeps}) is \sout{often} referred to as configuration interaction and the completeness of the configuration interaction representation is shown here by construction.

The computational cost of evaluating the configuration interaction representation can be analyzed similarly as in the symmetric case. The cost of evaluating a single Slater determinant $D_{\bm v}$ or its gradient $\nabla D_{\bm v}$ is of the order $O(N^3)$ (the dominant contribution is the factorisation of the $(\phi_{v_a}(i_b))_{ab}$). Since we are restricting the basis to $v_1 < v_2 <  \dots$ the number of terms can be bounded exactly as in the symmetric case; cf.~Eq.(\ref{eq:TD_bound}). Thus, for a total degree discretisation we obtain 
\[
    \COST \lesssim N^3 \cdot \PARAMS. 
\]
The strict ordering $v_1 < v_2 < \dots$ as opposed to non-strict ordering $v_1 \leq v_2 \leq \dots$ in the symmetric case only marginally improves the bound in the pre-asymptotic regime. 

\paragraph{Effective single determinant}
A complete expansion may also be obtained from a single determinant-like object by simply changing the summation over particles and basis function indices. Consider a general product expansion that is not symmetric with respect to exchange of particles,
\begin{equation} 
    \begin{split}
    &\varPhi(1 2 3 \dots N) = \\
    &\sum_{v_1 < v_2 < \dots < v_N  } c_{v_1 v_2 \dots v_N}  \phi_{v_1}(1) \phi_{v_2}(2) \dots \phi_{v_N}(N) \,, \label{eq:single}
    \end{split}
\end{equation}
then
\begin{equation}
    \label{anti-symmetrise_single}
    \Psi =  \frac{1}{\sqrt{N!}} \sum_{i_1 i_2 \dots i_N  }  \epsilon_{i_1 i_2 \dots i_N} \varPhi(i_1 i_2 \dots i_N)  \,
\end{equation}
is again a complete parameterisation of anti-symmetric functions. Despite having to anti-symmetrize only the function $\varPhi(1 2 3 \dots N)$, this representation appears computationally highly inefficient since the $O(N!)$ explicit anti-symmetrisation cannot be avoided as in the case of a Slater determinant. 
We will discuss the related backflow parameterizations in the following but already here hint that this representation my be viewed as an effective single Slater determinant by representing $\varPhi(1 2 3 \dots N)$ as a product of $N$ effective single-particle functions $\varphi_i$, where effective means that $\varphi_i$ depends on all particle coordinates, which allows one to achieve efficient representations.

\subsubsection{Restricted particle summation}

Alternatively, one may limit the sum over particle indices to $i_1 < i_2 < \dots < i_N$. As for $N$ particles for an anti-symmetric expansion there is only one contribution to the summation that fulfills $i_1 < i_2 < \dots < i_N$, the sum over particle indices may be dropped and the expansion written as
\begin{equation}
\Psi =  \sum_{v_1 v_2 \dots v_N  } c_{v_1 v_2 \dots v_N} \epsilon_{v_1 v_2 \dots v_N} \phi_{v_1}(1) \phi_{v_2}(2) \dots \phi_{v_N}(N) \,. \label{eq:asymACE}
\end{equation}
This corresponds to the occupation number representation of second quantization. Because of $\epsilon_{v_1 v_2 \dots v_N} $ two electrons cannot be in the same orbital and exchanging two electrons means exchanging their one particle orbitals which leads to a sign change of the expansion.

\section{Effective single particle functions}
\label{sec:effsinglepart}
In practise the configuration interaction and related representations need to be improved for accurate wave function representations. A basic idea is to regard the single electron wave functions $\phi_{v}(x)$ as quasi-particle wave functions that depend not only on the coordinate of particle $i$ but on the coordinates of other particles, too. This will also become a crucial ingredient in our review of nonlinear parameterisations below.

We use the ACE to construct a general representation of a generalized single particle function 
\[
    \varphi(i) = \varphi_i(x_i; {\bm x}_{\neq i})
\]
that is associated to particle $i$ but depends on all other particles, too. The notation $\varphi_i(x_i; {\bm x}_{\neq i})$ indicates that $\varphi_i$ is required to be symmetric with respect to exchange of all particles except $i$.
A complete expansion using the ACE formalism then reads
\begin{equation}
    \varphi(i) = \sum_{\bm v} c_{\bm v} \phi_{v_1}(i) \prod_{t = 2}^{\nu} A_{v_t} \,, \label{eq:effsingle}
\end{equation}
with
\begin{equation} \label{eq:effsingle_A_standard}
   A_{v} = \sum_{j \neq i} \phi_v(j) \,.  
\end{equation}

While Eq.(\ref{eq:effsingle}) provides a completely general expansion of an effective single-particle function, it still lacks two fundamental properties. First, the Hamiltonian is invariant under simultaneous translation of the positions of electrons and nuclei and the same should hold for the wave function, while the single electron basis functions $\phi_{v}(i)$ and the effective single electron functions $\varphi_i$ are not. A simple method to generate translationally invariant representations is to work with difference vectors between particles only. This was employed, for example, in ACE models for interatomic potentials\cite{Drautz19} to achieve translational invariance of general interatomic interactions. In addition the ACE representation provides a complete representation for functions with permutation symmetries. Secondly, the many-electron wave function has characteristic cusps when two electrons come close. It is computationally demanding to provide accurate representations of the cusps with smooth polynomial single electron basis functions. By working with basis functions that depend on difference vectors the cusps can be naturally injected into the wavefunction.

We therefore propose to replace the atomic base $A_v$ in Eq.(\ref{eq:effsingle_A_standard}) and $\phi_{v_1}(i)$ in Eq.(\ref{eq:effsingle}), respectively, with
\begin{align}
    \label{eq:A_with_cusps}
    A_{v} = \sum_j^{\text{electrons}} \phi_{v}^{\rm el}(\pmb{r}_j - \pmb{r}_i) + \sum_j^{\text{nuclei}} \phi_{v}^{\mu_j} (\pmb{R}_j - \pmb{r}_i)& \,, \\ 
    \text{and} \qquad \qquad \phi_{v}(i) = \sum_j^{\text{nuclei}} \varphi_{v}^{\mu_j} (\pmb{R}_j - \pmb{r}_i),&
\end{align}
where $\mu_j$ denotes the chemical species of atom $j$.
The linear combination of atomic orbitals appears to be a natural choice for representing the pairwise function,
\begin{align}
    \label{eq:R_with_cusps}
 \phi_v^{\rm el} (\pmb{r}) &= R_{nl}^{\rm el}(r) Y_l^m(\hat{\pmb{r}}) \,,\\
  \phi_{v}^{\mu_j} (\pmb{r}) &= R_{nl}^{\mu_j}(r) Y_l^m(\hat{\pmb{r}}) \,,
\end{align}
where $v = (nlm)$ is now a multi-index. Different from the usual reasoning, the orbitals are carried by the electrons and not the nuclei. In this representation it is also straightforward to integrate cusp conditions\cite{Fournais2005-uz} in the limit $\pmb{r}_{ji} \to 0$ through the choice of the radial basis $R_{nl}^{\rm el}, R_{nl}^{\mu_j}$.

The pairwise functions (or, equivalently, the atomic base) could be adapted,
\begin{equation} \label{eq:tucker}
    \begin{split}
    A^{\rm opt}_{v} &= \sum_k u_{vk} A_k, 
    \qquad \text{and} \\
    \phi_{v}^{\rm opt}(x_i) &= \sum_k u_{1,vk} \phi_{k}(x_i).
    \end{split}
\end{equation}
Replacing $A_v$ with $A_v^{\rm opt}$ and $\phi_{v_1}$ with $\phi_{v_1}^{\rm opt}$ in Eq.(\ref{eq:effsingle}) results in a nonlinear representation of $\varphi_i$ known as a Tucker tensor format. Such a format has been used with considerable success for optimizing only the radial basis in the original atomic cluster expansion\cite{Drautz19,lysogorskiy_performant_2021}.

The effective single particle functions could be extended and refined further by adapting recent extensions of ACE in natural ways to include message passing to electrons \cite{Bochkarev2022_multilayer,Batatia2022_Design} along similar lines as in the PauliNet archtecture\cite{Hermann2020-bf}.

For some applications covariance of the wave function and also the generalized single-particle functions $\varphi(i)$ under joint rotation and inversion of electron and nuclei positions may be required. Such covariance is readily incorporated into the parameterisation, using the techniques outlined in Refs.\onlinecite{Drautz19,Dusson20}, and results in 
\begin{equation}
    \varphi(i) = \sum_{k} c_k \sum_{{\bm v}} \mathcal{C}_{k, {\bm v}} \phi_{v_1}(i) \prod_{t = 2}^{\nu} A_{v_t},
\end{equation}
where the (sparse) matrix $\mathcal{C}$ contains generalized Clebsch--Gordan coefficients that specify all possible covariant couplings between the $\phi_{v_1}, A_{v_t}$. In particular, this symmetrisation results in significantly fewer parameters $c_k$ than parameters $c_{\bm v}$ in \eqref{eq:effsingle}.

\section{Nonlinear Parameterisations} 
\label{sec:nonlinear}
Since evaluating general anti-symmetric functions is  significantly more costly than evaluating general symmetric functions, there are numerous approaches to represent anti-symmetric functions in terms of symmetric contributions, normally leading to nonlinear parameterisations of anti-symmetric functions. In Eq.(\ref{eq:tucker}) we have already introduced a canonical mechanism how nonlinearities arise. Here we discuss how several other important approaches can be related to the atomic cluster expansion, or, how the atomic cluster expansion may be employed to represent both the symmetric and anti-symmetric contributions in these representations.

\subsection{Slater-Jastrow ansatz} 
Trivially, any anti-symmetric function $\Psi$ may be decomposed into 
\begin{equation} 
    \label{eq:generalprod}
    \Psi(1\cdots N) = \sum_{q = 1}^Q U_q(1\cdots N) \Psi_q(1\cdots N),
\end{equation}
where $U_q$ are symmetric and $\Psi_q$ again anti-symmetric. The challenge then is to impose restrictions on $\Psi_q$ that achieve desired accuracy or qualitative physical properties at an overall much lower cost than fully resolving the wave function directly using, e.g., the naive configuration interaction.

The original Jastrow (or, Slater--Jastrow) wave function ansatz\cite{Jastrow1955-cc, Foulkes2001-vi} takes the form 
\begin{equation} 
    \label{eq:jastrow_2b}
    \psi(1\dots N) = \exp\bigg( \sum_{i < j} u_2(r_{ij}) \bigg) \Psi_1(1 \cdots N),
\end{equation}
where $\Psi_1$ is an anti-symmetric function often chosen to be a classical Slater determinant. For example, if $\Psi_1$ is chosen to be a Slater determinant containing the standard atomic orbitals, and $u_2(r_{ij}) = a_{ij} r_{ij} / (1 + b_{ij} r_{ij})$ with suitably chosen parameters $a_{ij}, b_{ij}$ then the ansatz incorporates the correct electron-electron cusps. By incorporating the nuclei into the expression the electron-nuclei cusps can be resolved well. An in-depth review is given by \citet{Foulkes2001-vi}.

The atomic cluster expansion provides a computationally cheap and general mechanism to generalise the two-body form of the Jastrow factor to incorporate terms of arbitrary body-order, 
\begin{equation} \label{eq:jastrow_ace}
    \Psi(1 \dots N) = 
        \exp\bigg( \sum_{i = 1}^N \varphi^{(\nu)}(i) \bigg)
        \Psi_1(1 \dots N), 
\end{equation}
where $\varphi^{(\nu)}(i)$ is a generalized single particle function with ACE representation of the form Eq.(\ref{eq:effsingle}), possibly with restricted correlation-order $\nu$, and centered at electron $i$.

As discussed above, the computational cost associated with symmetric functions of higher correlation order is quite moderate compared to the cost of the configuration interaction approach. It is particularly interesting in this context to explore how the accuracy of this {\em ACE-Jastrow-Slater ansatz} improves with increasing correlation order. For example, whether higher correlation orders can be employed to resolve higher-order electron cusps.

\subsection{Vandermonde ansatz} 
The Slater-Jastrow ansatz (with a single term, i.e. $Q = 1$) has many desirable physical and computational properties but is in general understood to be incomplete. \citet{Han2019-bn} introduce a specific class of Slater determinants that further simplify it and render it a complete (universal) representation. If we choose $\Psi_q$ in Eq.(\ref{eq:generalprod}) to be a generalised Vandermonde determinant, 
\begin{equation}
    \Psi_q := \mathcal{V}({\bm \varphi}_q) := \prod_{i < j} (\varphi_q(i) - \varphi_q(j)), 
\end{equation}
where ${\bm \varphi}_q = (\varphi_q(i))_{i = 1}^N$ is a permutation covariant vector, and $U_q$ are {\em general} symmetric functions, then they establish that Eq.(\ref{eq:generalprod}), which now reads 
\begin{equation}
    \label{eq:vandermonde}
    \Psi = \sum_{q = 1}^Q U_q \mathcal{V}({\bm \varphi}_q),
\end{equation}
can be converged to a general anti-symmetric function. 

The ACE representation can obviously be used both to parameterise the symmetric part $U_q$, either directly or as a generalized Jastrow factor as in Eq.(\ref{eq:jastrow_ace}). In addition, ACE is also eminently suited to parameterise ${\bm \varphi}_q$; namely, a general permutation covariant vector can be written again as a generalized single particle function, 
\[
    \big[{\bm \varphi}_q\big]_i = \varphi_q(i)  
    = 
    \varphi_q\big(x_i; {\bm x}_{\neq i}\big),
\]
where $\varphi_q$ is symmetric in the second argument ${\bm x}_{\neq i}$ and hence represented in ACE format by Eq.(\ref{eq:effsingle}).
The results of \citet{Han2019-bn} imply that this representation is {\em complete} in the limit of an infinite basis and $Q \to \infty$. We are unaware of any attempt to employ this representation in practical simulations.

\subsection{Backflow ansatz} 
The backflow ansatz originally suggested by \citet{Feynman1956-hq} has increased in prominence due to the phenomenal success when incorporating deep learning strategies\cite{Pfau2020-zv,Hermann2020-bf}. It employs a  generalized Slater determinant, 
\begin{align}
    \label{eq:backflow}
    \Psi &= \mathcal{B}(\varphi_1, \dots, \varphi_N) \\ 
    \notag
    &=
    \det
    \begin{pmatrix}
        \varphi_1(1) & \varphi_2(1) & \cdots &  \varphi_N(1) \\
        \varphi_1(2) & \varphi_2(2) & \cdots & \varphi_N(2)\\
            \vdots & \vdots &   & \vdots \\
        \varphi_1(N) & \varphi_2(N) & \cdots & \varphi_N(N),
    \end{pmatrix}
\end{align}
where each $\varphi_j(i) = \varphi_j(x_i, {\bm x}_{\neq i})$ is permutation-invariant in the second argument. 
It is easy to see that this generalizes both the Slater-Jastrow ansatz Eq.(\ref{eq:jastrow_2b}) as well as the generalised Vandermonde representation Eq.(\ref{eq:vandermonde}).\cite{Huang2021-hb}
The backflow ansatz can be practically implemented using the ACE parameterisation of the components $\varphi_i$ by employing again the ACE parameterisations of Sec.~\ref{sec:effsinglepart}. Similarly as in the Slater--Jastrow approach, if ACE is employed to parameterise the $\varphi_j$ components, it becomes a both mathematically and physically interesting question to explore how the accuracy of the parameterisation improves with increasing correlation order.

\citet{Hutter2020-or} showed that a single term of this form suffices to represent a general anti-symmetric function. In particular, this establishes that also the ACE variant of Eq.(\ref{eq:backflow}) provides a complete parameterisation when correlation-order $\nu = N$ is chosen, and in the limit of an infinite basis. However, Hutter's proof says little about practical approximation rates since the the components $\varphi_j$ that he explicitly constructs are highly singular. 
Indeed, \citet{Huang2021-hb} showed by a dimensionality argument that general anti-symmetric polynomials $\Psi$ are extremely costly to represent in terms of the backflow parameterisation. This shows that there can be no general and simple argument to explain the success of the backflow (and by extension Slater--Jastrow) approaches. 

To conclude this discussion we make two elementary observations how the backflow ansatz can be obtained from linear parameterisations. The first observation is simply based on the standard Slater determinant, 
Eq.(\ref{eq:standard_slater_det}). If one were to give the orbitals $\phi_{v_1}, \dots \phi_{v_N}$ ``environment dependence'', i.e., they become many-particle functions 
$\phi_{v_i}(x_j; {\bm x}_{\neq j})$ then we immediately obtain the ACE variant of the backflow ansatz. In this perspective, one would trade the cost of multiple Slater determinants against the cost the orbitals $\varphi_j$.

The second observation starts from the computationally intractable formulation Eq.(\ref{anti-symmetrise_single}). Suppose, for the sake of argument, that we could decompose $\varPhi$ into a product
\begin{equation}  \label{eq:decomp_Phi_prod}
    \varPhi(1 2 3 \dots N) = \prod_{i=1}^N \varphi_{i}(x_i; {\bm x}_{\neq i})\,,
\end{equation}
where each $\varphi_i$ is symmetric with respect to exchange of particles except for $x_i$, then we directly obtain the backflow parameterisation, 
\begin{equation}
    \begin{split}
    \Psi &=  \frac{1}{\sqrt{N!}} \sum_{i_1 i_2 \dots i_N}^N \epsilon_{i_1 i_2 \dots i_N}  \prod_{t = 1}^N \varphi_{t}(x_{i_t}; {\bm x}_{\neq i_t})  \\
    &= 
    \mathcal{B}(\varphi_1, \dots, \varphi_N).
    \end{split} 
\end{equation}
Of course the decomposition Eq.(\ref{eq:decomp_Phi_prod}) is not general, hence this does not constitute a proof of completeness of the backflow ansatz, but a possible route towards identifying low-dimensional structures that could explain its significant practical success.

\section{Summary}
We demonstrate how the atomic cluster expansion naturally reproduces and relates the most common and most important representations of many-electron wave functions. It remains to be seen how this framework can be leveraged for the development of more efficient computational methods, or a deeper theoretical understanding of existing methods. Numerous routes exist that must be explored, such as the optimization of effective single particle functions to incorporate cusp conditions while retaining the regularity and low-dimensionality of sparse polynomial spaces, combined with a systematically improvable backflow ansatz.

For modelling atomic interactions, the polynomial ACE basis was shown to improve over machine learning representations. Analogous improvements for wave function representations may also be facilitated by ACE.

\begin{acknowledgements}
    RD acknowledges heplful discussions with Yury Lysogorskiy, Anton Bochkarev and Matteo Rinaldi. 

    CO thanks Huajie Chen, Dexuan Zhou and Zeno Sch\"{a}tzle for inspiring conversations on the topic of this article. 
    
    CO is supported by the Natural Sciences and Engineering Research Council of Canada (NSERC) [funding reference number IDGR019381].
\end{acknowledgements}

\end{document}